\documentclass[aip,jcp,reprint]{revtex4-1}

\usepackage{amsmath}
\usepackage{amssymb}
\usepackage{amsfonts}

\usepackage{graphicx}
\newcommand{\da}{^\dagger}

\newcommand{\ddx}[1]{{\partial_{#1}}}
\newcommand{\dddx}[1]{{\partial^2_{#1}}}
\newcommand{\dfdx}[2]{\dfrac{\partial#1}{\partial#2}}

\newcommand{\br}[1]{\langle #1 \vert}
\newcommand{\ke}[1]{\vert #1  \rangle}
\newcommand{\bk}[2]{\langle #1  \vert #2  \rangle}

\begin{document}

\title{Non-adiabatic dynamics of molecules in optical cavities}

\author{Markus Kowalewski}
\email{mkowalew@uci.edu}
\author{Kochise Bennett}
\author{Shaul Mukamel}
\email{smukamel@uci.edu}
\affiliation{Department of Chemistry, University of California, Irvine,
California 92697-2025, USA}
\date{\today}%

\begin{abstract}

Strong coupling of molecules to the vacuum field of micro cavities can modify the potential energy surfaces opening new photophysical and photochemical reaction pathways.
While the influence of laser fields is usually described in terms of classical field, coupling
to the vacuum state of a cavity has to be described in terms 
of  dressed photon-matter states (polaritons) which require quantized fields.
We present a derivation of the non-adiabatic couplings for single molecules in the strong coupling regime suitable for the calculation of the dressed state dynamics.
The formalism allows to use quantities readily accessible from quantum chemistry codes like the adiabatic potential energy surfaces and dipole moments to carry out wave packet simulations in the dressed basis.
The implications for photochemistry are demonstrated for a set of model systems representing typical situations found in molecules.
\end{abstract}

\maketitle

\section{Introduction}
The fate of a molecule after excitation with light is determined by its excited (bare) state potential
energy surface. While most molecules make their way back to the ground state by spontaneous
emission or non-radiative relaxation, some dissociate, isomerize, or are funneled through conical intersections (CoIn) \cite{Domcke}.
The reactivity can be manipulated either by chemical modification, changing the
environment, or by using photons to interact with the molecule while it evolves in an
excited states. 
It has been shown theoretically  \cite{Gollub10} and experimentally  \cite{Sussman06sci,Kim12jpca} that light can actively influence the molecular reactivity.
The modified photonic vacuum in nano scale fabricated cavities allows for
influencing molecular potential energy surfaces in a nondestructive manner and without the use
of external laser fields. Substantial couplings can be induced between electronic states with just a single photon \cite{Hutchison12ac}.
The radiation matter coupling  is enhanced  in small cavity modes \cite{Purcell46} and the strong coupling regime may be realized even when the field is in the vacuum state \cite{Morigi07prl,Kowalewski11pra}. The strongly coupled molecule+field states are known as dressed  atomic states \cite{Haroche} or polaritons \cite{Hopfield58pr}.
Recent experimental developments show promising results, paving the way for strong coupling
in the single molecule regime.
Strong coupling can be achieved in nano cavities \cite{Tomljenovic06oe}, nano plasmon antennas \cite{Kim15nl}, and nano guides \cite{Faez14prl}.
Chemical reactivity can be influenced in a very distinct way in this regime.
This provides great potential for the manipulation and control of e.g. the photo-stability of molecules, novel types of light induced CoIns, or modifying existing CoIns.
Specially tailored nano structured materials may then serve as a photonic catalysts that can be used instead of chemical catalysts.
In a recent theoretical paper Galego et al. \cite{Galego15prx} pointed out the impact of strong coupling on the absorption spectrum of molecules.

In the strong coupling regime the molecular and the photon degrees of freedom are heavily entangled and the molecular bare states do not provide a good basis. The quantization of the radiation field has to be
taken into account. This has been first described theoretically in the Jaynes-Cummings
model \cite{Jaynes63} which assumes an electronic two level system coupled to a single field mode and has been experimentally applied to atoms \cite{Hood98prl}
In molecules the nuclear degrees of freedom must be taken into account as well \cite{Galego15prx}.
The product basis of electronic and photonic states is not adequate
in the strong coupling regime. Diagonalizing the system to the dressed basis recovers
potential energy surfaces but also leads to light induced avoided crossing and actual curve crossings between the dressed states, analogous to avoided crossings and CoIns. The dynamics of the nuclei, electrons, and photons are strongly coupled in the vicinity of these crossings and pose formidable computational challenge.

Strong coupling to one or more radiation field modes can alter the molecular levels, profoundly  affecting the basic photophysical and photochemical  processes. The obvious way to achieve this regime is by subjecting the molecule to strong laser fields \cite{Kim12jpca,Kim15jpb,vdHoff11fd,Pritchard12arca}.
Alternatively the coupling can be enhanced by placing the molecule in a cavity and letting it interact with the localized cavity modes.
The coupling increases with $1/\sqrt{V}$, where $V$ is the mode volume.
Strong fields are not necessary in this case and the field can be even in the vacuum state. The former scenario can be realized
with classical fields.
This paper focuses on this latter, which involves quantum fields \cite{Jaynes63}.
A major difference between the two scenarios is the number of photons available in
the dressing field. A strong laser field can give rise to multiphoton absorption and
multiphoton ionization pathways that can interfere with the intended manipulation
of the quantum system.

We develop a formalism, which allows to express the dressed states and the non-adiabatic couplings in terms of readily accessible molecular properties like the bare state potential energy
surfaces and the transition dipole moments that can be extracted from standard  quantum chemistry calculations.
We demonstrate how chemical reactions can be modified by applying 
this theoretical framework to typical model systems.
We focus on a moderate coupling strength where the dressed state energies are not well separated but experience curve crossings giving rise to non-adiabatic dynamics.

The paper is structured as follows. In section \ref{sec:theory} we present the formalism
by including the nuclear degrees of freedom into the Jaynes-Cummings model. In section \ref{sec:Photochemistry} we present three models of molecular systems strongly coupled to the cavity. The photonic catalyst model couples a bound state to a dissociative
state, effectively opening a decay channel, decreasing its lifetime.
The photonic bound state model demonstrates how stimulated emission from the vacuum state
can increase the lifetime of a otherwise unbound state.
Finally, forming light induced conical intersections in a cavity mode is demonstrated on the formaldehyde molecule.

\section{Theoretical Framework}
\label{sec:theory}
We use the Jaynes-Cummings (JC) model \cite{Jaynes63} to describe the coupling of the resonator to the molecular dipole transition.
\begin{equation}
H_{JC} = H_M + H_C + H_I\,,
\end{equation}
where $H_M$ is the molecular Hamiltonian representing two electronic states
\begin{align}
H_M = \frac{\hbar}{2} \omega_0 \left( \sigma\da\sigma- \sigma\sigma\da \right)\,,
\end{align}
$H_C$ is cavity Hamiltonian of a single quantized photon mode
\begin{align}
H_C = \hbar\omega_c \left( a\da a + \frac{1}{2}\right)\,,
\end{align}
and $H_I$ describes the interaction between
the photon mode and the molecule
\begin{eqnarray}
H_I = \hbar g \left(a\da\sigma+a\sigma\da \right)\,.
\end{eqnarray}
Here $\sigma\da = \ke{e}\br{g}$ and $\sigma = \ke{g}\br{e}$ are the creation and annihilation of a molecular excitation of the molecular eigenstates in the electronic subspace $\ke{g}$ and $\ke{e}$. The excitation energy between the bare eigenvalues $\omega_g$ and $\omega_e$ is $\omega_0=\omega_e - \omega_g$.
The cavity mode with frequency $\omega_c$ is described by the eigenstates $\ke{n_c} \equiv \ke{0}$, $\ke{1}$, ... .
$a\da$ and $a$ are the bosonic creation and annihilation operators of the cavity mode.
The interaction $H_I$ is given in the rotating wave approximation (RWA), where $g=\varepsilon_c \mu_{eg}/2\hbar$ is the coupling strength. The RWA holds
when $\delta_c \ll \omega_0$ and $g \ll \omega_0$.
 $\mu_{eg}$ is the molecular transition dipole moment and
$\varepsilon_c$ is the cavity vacuum field,
\begin{equation}
\label{eq:evac}
\varepsilon_c = \sqrt{\frac{\hbar \omega_c}{V\epsilon_0}}\,,
\end{equation}
where $V$ is the resonator mode volume.

The eigenstates of $H_{JC}$ are the dressed (polariton) states $\ke{\pm,n_c}$:
\begin{eqnarray}
\label{eq:dressedstates}
\ke{+,n_c} =&  \cos{\theta} \ke{e,n_c} + \sin{\theta} \ke{g,n_c+1}\\
\ke{-,n_c} =& -\sin{\theta} \ke{e,n_c} + \cos{\theta} \ke{g,n_c+1}\,,
\end{eqnarray}
where the mixing angle $\theta$ is
\begin{equation}
\label{eq:mixangle}
\cos{\theta} = \sqrt{\dfrac{\Omega_n-\delta_c}{2\Omega_n}}, ~~~
\sin{\theta} = \sqrt{\dfrac{\Omega_n+\delta_c}{2\Omega_n}}
\end{equation}
with the corresponding eigenvalues
\begin{equation}
\label{eq:dse}
E_{\pm,n} = \dfrac{\hbar}{2} \omega_0 + \hbar \omega_c \left(n_c + \dfrac{1}{2}\right) \pm \dfrac{\hbar}{2} \Omega_n \,,
\end{equation}
and $\Omega_n$ is the Rabi-frequency
\begin{equation}
\Omega_n = \sqrt{4g^2 (n_c + 1) + \delta_c^2}\,.
\end{equation}
The molecule-cavity detuning
\begin{align}
\delta_c = \omega_0 - \omega_c = (V_e - V_g)/\hbar - \omega_c
\end{align} 
represents the frequency miss-match between the molecular transition and the cavity mode.
Here $E_e$ and $E_g$ are the eigenvalues of the bares states.
We assume that the cavity is initially in the vacuum state (i.e. $n_c=0$) and
omit the photon number $n_c$ in the following.

\subsection{The molecular Hamiltonian in the strong coupling regime}
The original JC model was developed for atomic transitions and does not
include nuclear degrees of freedom.
The molecular potential energy surfaces become coupled when
the electronic ground and excited state get into resonance with the cavity mode.
The nuclear and electronic motions will then be coupled and the Born-Oppenheimer approximation
breaks down.

To obtain the couplings in the dressed state basis we include the dependence
of the nuclear coordinates $q=(q_1, \dots, q_N)$ into the JC model. The
quantities $\delta_c$, $\Omega_n$, $g$ depend parametrically on $q$, and
the mixing angle $\theta$ (Eq. \ref{eq:mixangle}) also becomes a function of the nuclear coordinates.
The new dressed potential energy surfaces can then be expressed in terms of the dressed state eigenvalues
of Eq. \ref{eq:dse}:
\begin{align}
V_{\pm,0} &= \dfrac{1}{2} (V_e+V_g) \pm \dfrac{\hbar}{2} \Omega_0\\
V_{g,0} &= V_g
\label{eq:Vpm}
\end{align}
where $V_g \equiv V_g(q)$ and $V_e \equiv V_e(q)$ are the bare state potential energy surfaces of
the free molecule.

We follow the standard procedure to derive the non-adiabatic coupling terms
in the adiabatic basis \cite{Domcke,Hofmann01cpl,Martinez97jpca}.
Atomic units are used in the following  ($\hbar = m_e = 4\pi\epsilon_0 = 1$).
Instead of the bare adiabatic electronic states, we use the dressed states from Eqs. \ref{eq:dressedstates} denoted $\ke{\phi_k}\equiv \ke{\phi_k(r,q)}$, where $r=(r_1, \dots, r_M)$ are the electronic coordinates.
The total wave function is expanded in the adiabatic basis:
\begin{align}
\Psi=\sum_k \psi_k(q) \phi_k(r,q)
\end{align}
where $k$ runs over the set of dressed states ($\ke{g,0}, \ke{\pm,0}, \ke{e,1}$).
The full molecular Hamiltonian
\begin{align}
\hat H = \hat T + \hat H_{el}
\end{align}
consists of the nuclear kinetic energy term
\begin{align}
\hat T = -\sum_i\dfrac{1}{2m_i} \dfdx{^2}{q_i^2}
\end{align}
and the electronic part $\hat H_{el}$ with the parametric
eigenvalues $V_g(q)$ and $V_e(q)$. Taking the matrix elements $\br{\Psi} \hat H \ke{\Psi}$
and integrating over $r$ yields:
\begin{align}
\label{eq:Hkl}
\hat{H_{kl}} = \hat T +\delta_{kl} \hat V_{kl} +  \sum_i \dfrac{1}{m_i} \left( f_{kl}^{(i)} \dfdx{}{q_i} + \dfrac{1}{2} h_{kl}^{(i)} \right)
\end{align}
where $f$ and $h$ recover the derivative coupling term and the scalar coupling as they appear
in the theory of CoIns:
\begin{align}
f_{kl}^{(i)}(q) &= \br{\phi_k(q)} \ddx{q_i} \ke{\phi_l(q)}_r\label{eq:f}\\
h_{kl}^{(i)}(q) &= \br{\phi_k(q)} \dddx{q_i^2} \ke{\phi_l(q)}_r
\end{align}
No assumptions have been made on the bare electronic states. This result holds even
if $V_g$ and $V_e$ undergo a crossing. In the following we discuss the relevant matrix elements
of $f$ and $h$ in the dressed states basis and show how the cavity affects the non-adiabatic couplings.

Inserting the definitions of the dressed states (Eqs. \ref{eq:dressedstates}) into Eq. \ref{eq:f} yields the derivative coupling term between $\ke{-,0}$ and $\ke{+,0}$:
\begin{align}
\label{eq:pmdq}
f_{-,+}^{(i)}= \dfrac{\Delta G_i}{4g}
\left(1 - \dfrac{\delta_c^2}{4g^2+\delta_c^2} \right) - \dfrac{\delta_c}{4g^2+\delta_c^2}\dfdx{g}{q_i}
\end{align}
where $\Delta G_i = \ddx{q_i} (V_e-V_g)$ is the gradient difference.
The dressed state coupling has two contributions: The first term is governed by
the gradient difference of the two bare states PESs, whereas the second term depends on the gradient of the transition dipole moment through $\ddx{q_i}g$.
The latter vanishes in the Condon approximation but may be substantial in regions where the transition dipole varies rapidly with $q$.
Note that Eq. \ref{eq:pmdq} does not contain any coupling terms involving the bare state crossings ($f_{g,e}^{(i)}$) since these couplings vanish due to the orthogonality of the photon states.
This is in contrast to the couplings between the ground and the dressed states which solely contain the bare state derivative couplings but no contribution from the cavity:
\begin{align}
f_{g,+}^{(i)} &= f_{g,e}^{(i)}\cos \theta \\
f_{g,-}^{(i)} &= -f_{g,e}^{(i)}\sin \theta 
\end{align}
These terms may be safely neglected when the bare state energies are well separated. 
Note that all diagonal matrix elements of $f$ vanish ($f_{kk}=0$).

To evaluate the scalar coupling terms $h$ of the second derivatives we introduce the following
decomposition, which breaks down the equations and simplifies the results.
\begin{align}
h_{kl}^{(i)} = \ddx{q_i} f_{kl}^{(i)} - F_{k,l}^{(i)}
\end{align}
The second term $F_{k,l}^{(i)} =  \bk{\ddx{q_i}\phi_k}{\ddx{q_i}\phi_l}$ now contains also diagonal contributions:
\begin{align}
\label{eq:Fpp}
F_{+,+}^{(i)} &= F_{g,g}^{(i)}\sin^2 \theta + F_{e,e}^{(i)} \cos^2 \theta
+ \dfrac{\Lambda_i^2}{4} + \dfrac{ \delta_c^2\Lambda_i^2}{16 g^2}\\
F_{-,-}^{(i)} &= F_{g,g}^{(i)} \cos^2\theta + F_{e,e}^{(i)} \sin^2 \theta
+ \dfrac{\Lambda_i^2}{4} + \dfrac{ \delta_c^2\Lambda_i^2}{16 g^2}\\
F_{-,+}^{(i)} &= \sin\theta\cos\theta \left( F_{g,g}^{(i)} - F_{e,e}^{(i)} \right)\\
F_{g,+}^{(i)} &= F_{g,e}^{(i)} \cos\theta + \dfrac{\Lambda_i f_{ge}^{(i)}}{4\cos\theta}\\
F_{g,-}^{(i)} &= -F_{g,e}^{(i)}\sin\theta + \dfrac{\Lambda_i f_{ge}^{(i)}}{4\sin\theta}\label{eq:Fg-}
\end{align}
with
\begin{align}
\Lambda_i =  \dfrac{\delta_c}{\Omega^3}
\left(4g\dfdx{g}{q_i}+\delta_c \Delta G_i \right) - \dfrac{\Delta G_i}{\Omega}
\end{align}
$f_{kl}^{(i)}$ and $F_{kl}^{(i)}$ contain all possible couplings:
intrinsic non-adiabatic couplings  of the bare states and cavity induced non-adiabatic couplings.
The Hamiltonian eq. \ref{eq:Hkl} thus describes the dynamics in the most general case.
The only approximation made is the RWA and the condition that the system
can not access higher photon states during the time evolution.
For very large detunings $\delta_c$ higher photon states ($n_c > 1$) must be taken into account.

The non-adiabatic couplings may be further simplified in specific parameter regimes.
Assuming that the bare states are well separated in energy and do not undergo any curve crossings,
all terms $f_{g,e}^{(i)}$, $F_{g,g}^{(i)}$, $F_{e,e}^{(i)}$, and $F_{g,e}^{(i)}$ may be neglected. The $F_{g,e}^{(i)}$ terms are usually neglected in molecular dynamics simulations and quantum dynamics of the bare states. Note that $F_{-,+}$ does not contain any contribution from the cavity.
$F_{\pm,\pm}^{(i)}$ vanishes for small gradient
differences and in the Condon approximation and may also be neglected in most
cases, since they only make a minor contribution to the shape of the PESs.
Dropping all $F$ terms leads to the approximate Hamiltonian:
\begin{align}
\label{eq:Hkl_simple}
\hat{H_{kl}} = \hat T +\delta_{kl} \hat V_{kl} +  \sum_i \dfrac{1}{2m_i} \left( 2f_{kl}^{(i)} \dfdx{}{q_i} +  \dfdx{}{q_i}  f_{kl}^{(i)} \right)
\end{align}
The hermitian Hamilton operator (Eq. \ref{eq:Hkl_simple}) will be used in the following
to calculate the wave packet dynamics.
Hamiltonians with this structure are commonly used to simulate the dynamics
in the vicinity of Conical intersections \cite{Hofmann01cpl} by numerical propagation of the
wave function.
This is done by using a grid in the nuclear coordinates,
rather than expanding in nuclear eigenstates which scales unfavorably with the number of nuclear modes.
Hereafter we use this approach.

Operators which represent molecular properties can be expressed in the bare state basis by transforming them into the dressed state basis using Eqs. \ref{eq:dressedstates} to \ref{eq:mixangle}.
The transition dipole moments then read:
\begin{eqnarray}
\label{eq:dipconv1}
\br{+,0}\hat\mu\ke{g,0} &=& \cos{\theta} \mu_{eg}\\
\label{eq:dipconv2}
\br{-,0}\hat\mu\ke{g,0} &=& - \sin{\theta} \mu_{eg} \\
\label{eq:trdipds}
\br{+,0}\hat\mu\ke{-,0} &=& \cos{\theta}\sin{\theta} \left( \mu_{ee} - \mu_{gg} \right)\,,
\end{eqnarray}
where $\mu_{eg}$ is the bare state electronic transition dipole moment and $\mu_{gg}$ and
$\mu_{ee}$ are dipole moments of the ground and excited state respectively. 

\section{Photochemistry in the Strong Coupling Regime}
\label{sec:Photochemistry}
\begin{figure}
\includegraphics[width=0.5\textwidth]{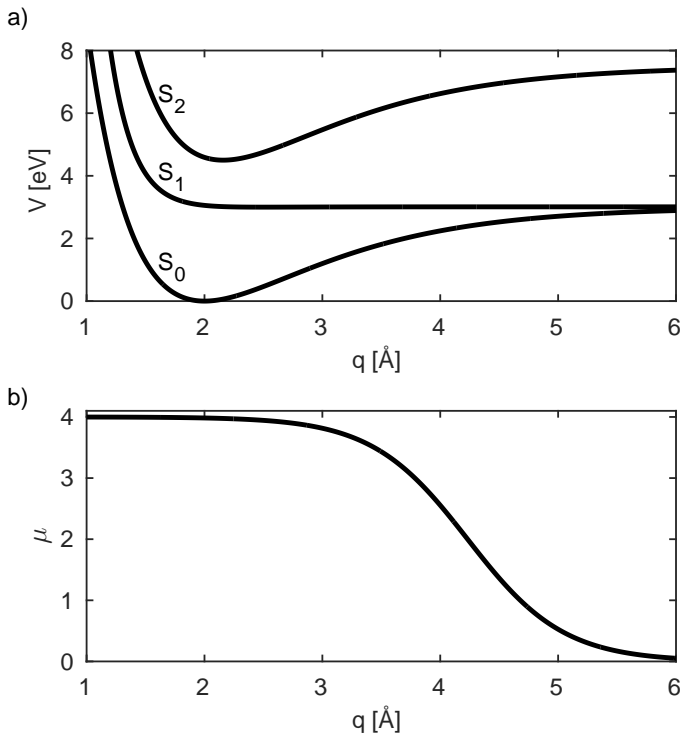}
\caption{(a) Bare state PESs used for the photonic catalyst model as well as the photonic bound state model. The minimum of the S$_2$ state is displaced by 0.3\,$\mathrm{\AA}$ with respect to the ground state. (b) Transition dipole curve used in the different models.
The parameters for the model are given in Appendix \ref{app:params}.}
\label{fig:pcat_bare}
\end{figure}
In the following we present calculations on three simple model systems to illustrate the basic possibilities
of the cavity coupling and the effects on the non-adiabatic couplings. The level structure of
the dressed states creates new pathways for the nuclear dynamics and new transitions for
spectroscopic measurements.
Our goal is to use the influence of the cavity to modify the reactivity of a molecule. 
Photodissociation in the dressed state basis can then be enhanced or suppressed.

\subsection{Photonic Catalyst}
\begin{figure}
\includegraphics[width=.5\textwidth]{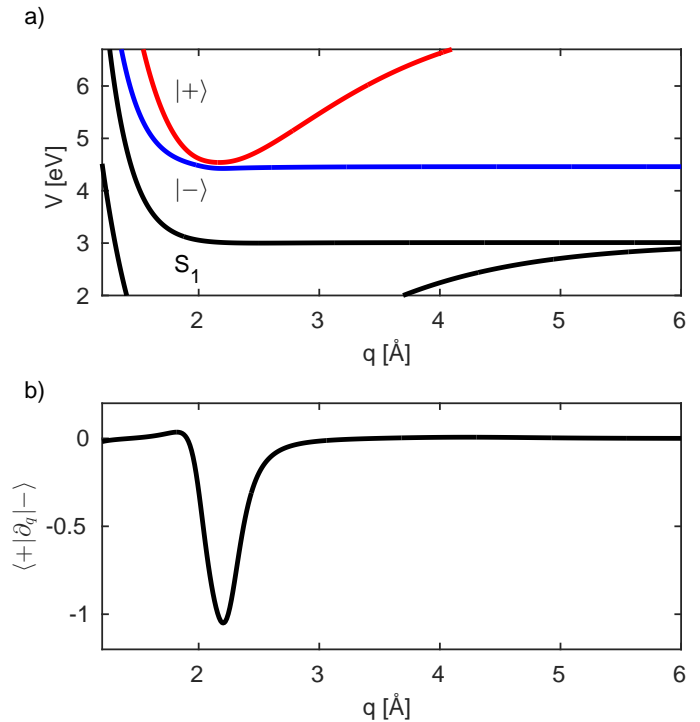}
\caption{(a) Dressed state PESs for the photonic catalyst model. The $S_2$ state is now
coupled to the dissociative state. (b) Non-adiabatic coupling matrix element for the polariton states.
Dominated by the gradient difference term of eq. \ref{eq:pmdq}. Parameters: $g=54$\,meV}
\label{fig:pcat_JC}
\end{figure}
In the first model (Fig. \ref{fig:pcat_bare}(a)) we assume that a bound state S$_2$ is accessible by a dipole
transition from the ground state S$_0$.
The dissociative state S$_1$ does not have a transition dipole moment with the ground state, but is accessible from S$_2$.
The cavity couples the states S$_2$ and S$_1$ through the transition dipole moment shown
in Fig. \ref{fig:pcat_bare}(b).
The cavity mode frequency $\omega_c$ is set be in resonance at the minimum of S$_2$ (1.45\,eV) and with a maximum cavity coupling of $g=54$\,meV.
The states $\ke{g}\equiv \mathrm{S}_1$ and $\ke{e}\equiv \mathrm{S}_2$ are used along with Eq. \ref{eq:Vpm} to form the dressed states, shown in Fig. \ref{fig:pcat_JC}(a).
The resulting dressed states $\ke{\mathrm{S}_1}$, $\ke{-}$, and $\ke{+}$ undergo
an avoided crossing close to resonance, while their shape remains similar to the bare states.
The corresponding non-adiabatic coupling matrix element $f_{+,-}$ (Eq. \ref{eq:f}), which is responsible for the transition between the dressed states is shown in Fig. \ref{fig:pcat_JC}(b).
The initially dark state S$_1$ now becomes radiatively accessible from S$_0$ through the non-adiabatic couplings via the S$_2$ state.
It is evident that the dressed states are coupled to each other in the region where the bare states are close to resonance with the cavity mode.
The upper dressed state -- whose shape still resembles the shape of the S$_2$ state -- is thus not stable anymore and the molecule can dissociate through the non-adiabatic coupling to the unbound lower dressed state.

\begin{figure}
\includegraphics[width=0.5\textwidth]{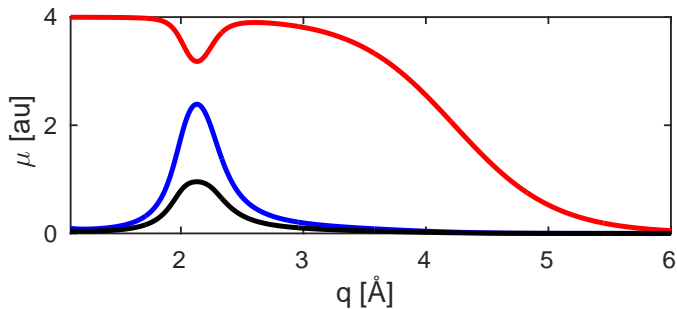}
\caption{Transition dipole moments in the dressed state basis for the photonic catalyst model:
$\mu_{g+}$ (red), $\mu_{g-}$ (blue),$\mu_{-+}$ (black).}
\label{fig:pcat_mu}
\end{figure}
Figure \ref{fig:pcat_mu} displays the relevant transformed transition dipole moments
calculated from Eqs. \ref{eq:dipconv1} to \ref{eq:trdipds}.
All curves show a dip around 2.2\,$\mathrm{\AA}$ where the non-adiabatic coupling and thus the mixing between the molecular states and the photon states is strongest.
The transition dipole between the dressed states vanishes if the cavity coupling vanishes.
The coupling to the cavity creates a new transition and modified dynamics
which can be probed with time resolved spectroscopy.

\begin{figure}
\includegraphics[width=0.5\textwidth]{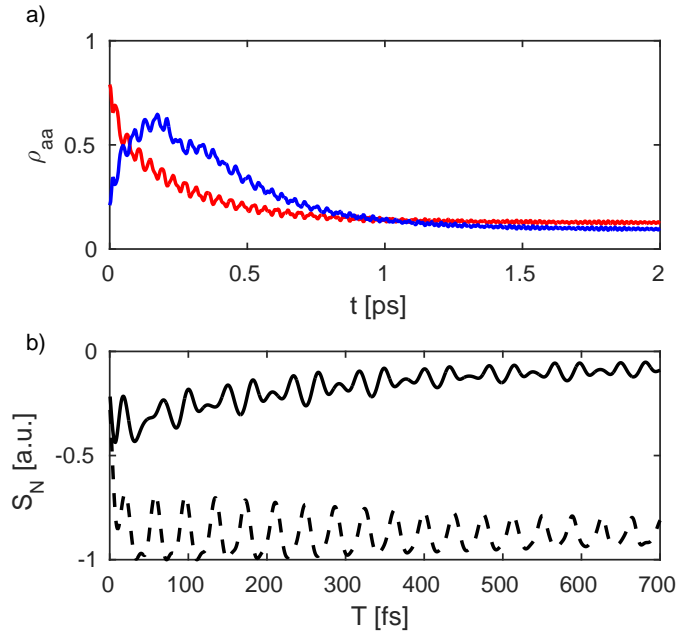}
\caption{(a) Populations of the dressed states $\ke{+}$ (red) and $\ke{-}$ (blue) in the photonic catalyst model vs. time. The decay of the  $\ke{+}$ state can be fitted to a bi-exponential model
yielding the time constants 228\,fs and 42\,ps.
(b) Transient absorption signal in depenence of the probe delay $T$ and. The laser is set to be resonant between S$_1$ and the $\ke{\pm}$ state (1.5\,eV) and has a pulse length of 10\,fs (FWHM).
The dashed line is the signal for a wave packet in the S$_2$ bare state potential.}
\label{fig:pcat_sig}
\end{figure}
The excited state evolution was simulated by wave packet dynamics on a spatial grid
(for details see Appendix \ref{app:Propgation}).
Figure  \ref{fig:pcat_sig}(a) depicts the dynamics after impulsive excitation from the ground state (S$_0$)
to the dressed states $\ke{\pm}$.
The initial population pattern is caused by the mixing of the transition dipole moments, followed by a rapid decay of the upper dressed state caused by the dissociative/unbound character of the lower dressed state. The oscillation pattern is caused by the wave packet oscillation in the $\ke{+}$ state,
passing through the coupling region.
Figure \ref{fig:pcat_sig}(b) shows the transient absorption signal (see Appendix \ref{app:signal})  probing the system via the $\ke{g}$ state. The signal shows a clear decay of stimulated emission modulated by the wave packet motion in the $\ke{+}$ state.

\subsection{Photonic Bound States}
\begin{figure}
\includegraphics[width=0.5\textwidth]{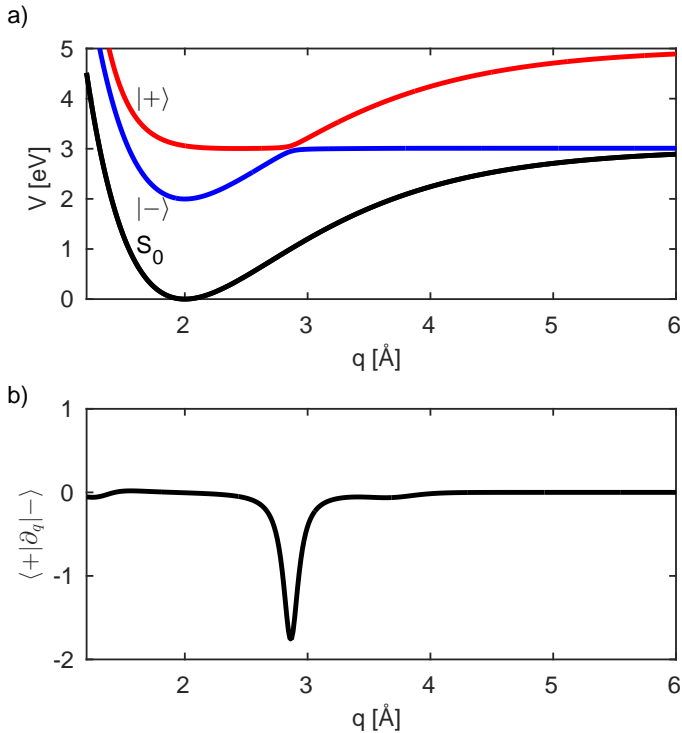}
\caption{(a) Dressed state PESs for the photonic bound state model.
(b) Non-adiabatic coupling matrix element causing transition between the dressed states.}
\label{fig:pem_JC}
\end{figure}
In the second model, we reverse the roles of bound and unbound states to create a situation where a purely
dissociative state can be stabilized (i.e. increase its lifetime) via cavity coupling to a bound state.
We use the model from Fig. \ref{fig:pcat_bare}, but only considering states S$_0$ and S$_1$. An excitation from S$_0$ to S$_1$ causes immediate dissociation of the molecule
in the bare state model.
Setting the cavity mode on resonance with S$_0$ and S$_1$ at $\approx 2$\,eV creates a set of dressed states, which experience an avoided crossing (Fig. \ref{fig:pem_JC}(a)) with a non-adiabatic coupling matrix element (Fig. \ref{fig:pem_JC}(b)) peaking at the crossing at 2.9\,$\mathrm{\AA}$.
The lower dressed state resembles the ground state around the Franck-Condon point and forms
a bound state potential. The upper dressed state now also appears as a partially bound state potential, which is coupled to the dissociative curve by the avoided crossing.

\begin{figure}
\includegraphics[width=0.5\textwidth]{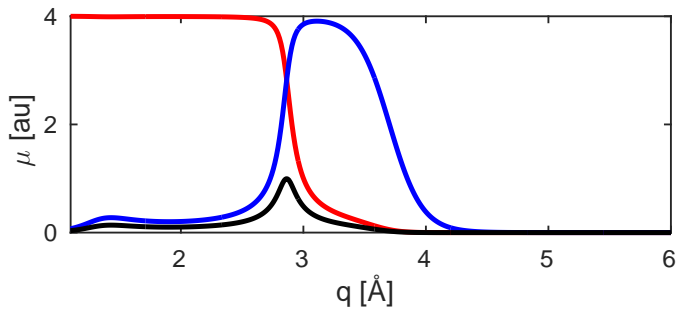}
\caption{Tranistion dipole moments in the dressed state basis: $\mu_{g+}$ (red), $\mu_{g-}$ (blue), $\mu_{-+}$ (black).}
\label{fig:pem_mu}
\end{figure}
The transition dipole moments are shown in Fig. \ref{fig:pem_mu}.
Due to the large detuning $\delta_c$ in the Franck-Condon region, the lower dressed state has a weak transition dipole moment with respect to the S$_0$ state. The state character change at the crossing at 2.9\,$\mathrm{\AA}$ manifests itself in the rapid change of the transition dipole moments (the crossing of the red and blue curve in Fig. \ref{fig:pem_mu}.).

\begin{figure}
\includegraphics[width=0.5\textwidth]{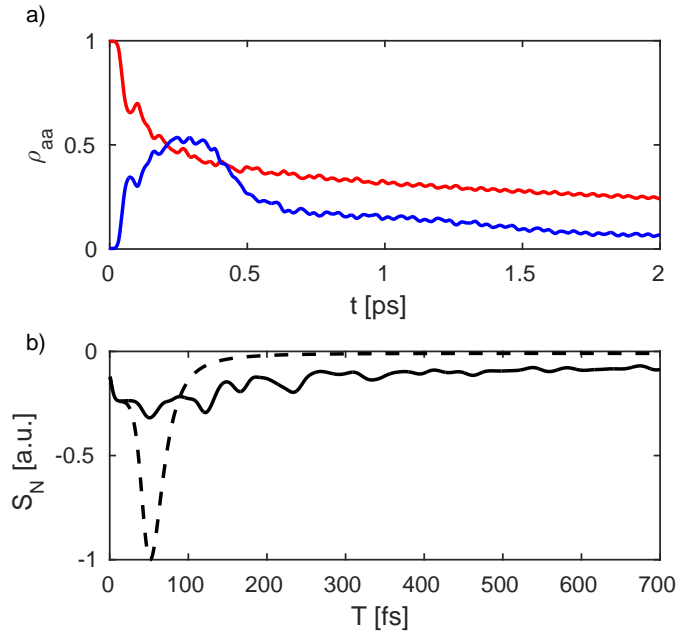}
\caption{(a) Population of the dressed states $\ke{+}$ (red) and $\ke{-}$ (blue) in the photonic catalyst model vs. time. The decay of the  $\ke{+}$ state can be fitted to a bi-exponential model
yielding the time constants 234\,fs and 5.2\,ps.
(b) Transient absorption signal. Laser is set to be resonant between S$_1$ and the $\ke{\pm}$ state (1.5\,eV) and has a pulse length of 10\,fs (FWHM).
The dashed line is the signal for a wave packet in the S$_1$ bare state potential}
\label{fig:pem_sig}
\end{figure}
The population dynamics after excitation is shown in Fig. \ref{fig:pem_sig}(a)).
The clear distribution of the dipole moments between ground state and the dressed states
leads to the upper dressed state population upon impulsive excitation.
The quasi-bound character of the upper dressed state becomes clear from Fig. \ref{fig:pem_sig}(a)):
Instead of immediate dissociation the upper dressed state acquires a significant lifetime. The
population of $\ke{+}$ leaks into $\ke{-}$ on a picosecond time scale. The corresponding
transient absorption signal is shown in Fig. \ref{fig:pem_sig}(b)) along with the signal
for the bare state system (dashed curve).

\subsection{Photoninduced Conical Intersections}
\begin{figure}
\includegraphics[width=0.5\textwidth]{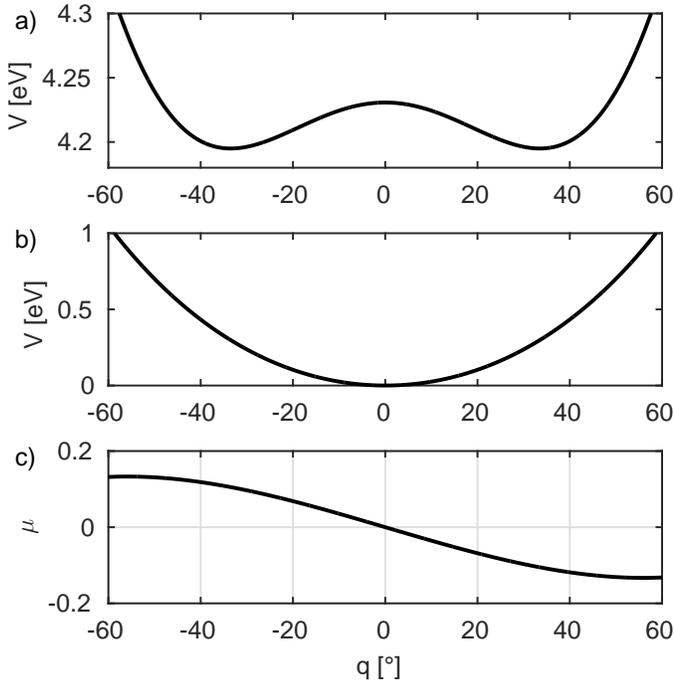}
\caption{Bare state PESs for the photoninduced CoIn model
calculated at the CAS(4/4)/MRCI/6-311G* level of theory with the program
package MOLPRO \cite{MOLPRO_brief}:
(a) The double minimum of the $S_1$ state.
(b) The ground $S_0$ ground state of formaldehyde. $q$ is the angle of the out-of-plane motion.
The cavity is set in resonance at $q=0$.
(c) Transition dipole moment $S_0 \rightarrow S_1$.}
\label{fig:phCI_bare}
\end{figure}
\begin{figure}
\includegraphics[width=0.5\textwidth]{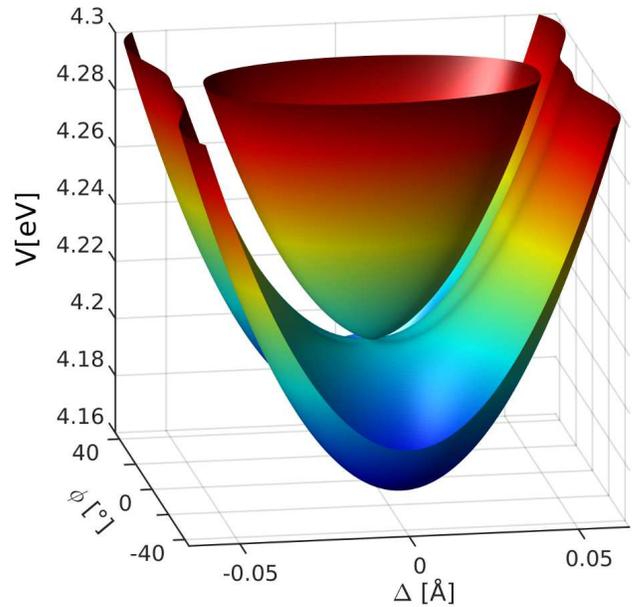}
\caption{Photoninduced CoIn  between the dressed states in formaldehyde in dependence of two vibrational modes: CH$_2$ out-of-plane motion ($\phi \equiv q_1$) and the  CH$_2$ asymmetric stretch motion ($\Delta \equiv q_2$). }
\label{fig:phCI_ds_2D}
\end{figure}

So far we have demonstrated the non-adiabatic couplings induced by the cavity in terms
avoided curve crossings, which stem from the fact that a non-vanishing dipole in the coupling
region creates a splitting between the dressed states (see Eq. \ref{eq:dse}). 
However, by choosing a point of vanishing transition dipole moments one can in principle also
create a crossing, which exhibits a degeneracy between the dressed states.
This is the basic requirement to obtain a CoIn, i.e.
the coupling between the adiabatic electronic states has to vanish at the intersection
point \cite{Domcke}.
In our third example of light-induced CoIns \cite{Demekhin13jcp,Kim15jpb} this condition is fulfilled by rotating the
molecule with respect to the polarization vector of the driving field.
By inspecting Eq. \ref{eq:Vpm} we identify another type of light induced CoIn:
Setting the cavity on resonance at a nuclear configuration where the transition dipole moments vanishes yields a degenerate point in the dressed state basis.
This can be achieved by choosing an electronic transition which is dipole forbidden at a certain
configuration of high symmetry and becomes allowed as the symmetry is lowered.
We now demonstrate this case for formaldehyde.
In its planar equilibrium
structure the lowest energy transition from the $^1A_1$ state to the $^1A_2$ state transition is dipole forbidden. Every vibrational mode which is not of the $A_1$ irreducible representation
breaks the $C_{2v}$ symmetry ($B_1$, $B_2$) and can be expected to make the transition dipole allowed.
In Fig. \ref{fig:phCI_bare} the potentials and transition dipole moments are shown vs. the out-of-plane motion ($B_1$) of the hydrogen atoms. 
Setting the cavity in resonance with the forbidden transition thus creates a vacuum field, light induced CoIn, which we call photoninduced CoIn.
\begin{figure}
\includegraphics[width=0.5\textwidth]{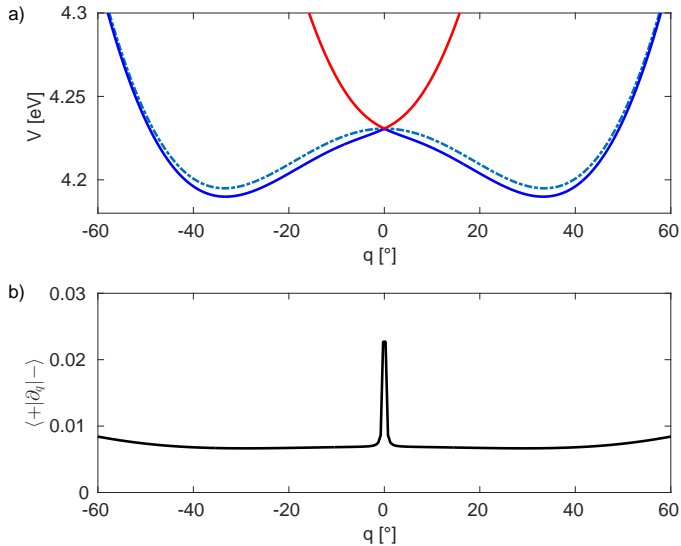}
\caption{(a) Dressed state PESs (red: $\ke{+}$, blue $\ke{-}$) for the photoninduced CoIn model in dependence of the out-of-plane angle $\phi$.
The dashed line indicates the $S_1$ bare state. (b) Non-adiabatic coupling matrix element for the polariton states. The splitting vanishes at $q=0$, hence the degeneracy.
The cavity coupling is chosen to be $g_{max}=434$\,meV.}
\label{fig:phCI_ds}
\end{figure}
The corresponding dressed states and the non-adiabatic coupling matrix element are shown
in Fig. \ref{fig:phCI_ds}. The degenerate point appears at the planar configuration ($q_1=0$)
along with a peaking non-adiabatic transition matrix element $f^{(1)}_{-,+}$. Note that
$f^{(1)}_{-,+}$ diverges when the detuning is exactly zero. 
Choosing a second vibrational mode which also breaks the symmetry will
create a transition dipole moment and will result in a typical cone-shaped PES.
We demonstrate this feature for the asymmetric stretch motion of the CH$_2$ group ($B_2$).
The resulting PES of the dressed states is shown in Fig. \ref{fig:phCI_ds_2D}.

\section{Conclusions and Outlook}
We have developed a theory,
which can be applied to compute the non-adiabatic dynamics of single molecules strongly coupled to a single cavity mode.
The formalism expressed in terms well known derivative
couplings is suitable for various simulations protocols like full quantum propagation
and semi-classical methods like for example surface hopping \cite{HammesSchiffer94} and ab initio multiple
spawning \cite{BenNun00jpca}. The quantities required to express the molecular system in terms of dressed states, i.e. the potential energy surfaces and transition dipole moments can be directly obtained from state of the art quantum chemistry methods.
The derivations are done within the RWA and the assumption that the
ultrafast dynamics timescale is shorter than the lifetime of the photon mode.
However, we could identify situations where
the JC Hamiltonian is not adequately described by two photon states and higher
manifolds must be taken into account for the procedure to converge.
A break down of the RWA can be caused by various factors: Large detunings give rise to off-resonant
terms ($a^\dagger\sigma^\dagger$ and $a\sigma$) in the ultra strong coupling regime
($g\approx \omega_c$) give rise to the Bloch-Siegert shift and ground state modifications.
The off-resonant regime can be easily accessed by coupling close to the CoIn, while the 
ultra strong coupling regime might be difficult to reach due to technical limitations
of the nano-structures, like for example the achievable size of the mode volume.
Moreover, the applicable field strength is limited by the ionization potential of a molecule.

Some basic possibilities for the manipulation of the excited state photo chemistry have been
demonstrated for photo dissociation model systems. The life time with respect
to dissociation can potentially be significantly influenced by the cavity coupling. Non-adiabatic
coupling between the dressed states is the cause for the coupling and the effect on the nuclear
dynamics. The population transfer between the dressed states may also be viewed
via stimulated emission caused by the vacuum state of the photon mode.

Single molecule strong coupling is an experimentally challenging regime and has not been demonstrates yet to the extent necessary to influence chemistry. However, coupling of
a ensemble of $N$ molecules to the mode of a micro resonator, which is enhanced by a factor $\sqrt{N}$ \cite{Hutchison12ac} shows promising
results. The collective chemistry in a cavity is a many body effect, which needs
further investigation. Its theoretical treatment is more challenging since all
particles are coupled and the dimensionality increase with the number of particles.
Finally the super radiant \cite{Dicke54pr} regime might be used to engineer the reactivity of molecules in a novel way.

\begin{acknowledgments}
The support  from the National Science Foundation (grant CHE-1361516) and support of the Chemical Sciences, Geosciences, and Biosciences division, Office of Basic Energy Sciences, Office of Science, U.S. Department of Energy through award No. DE-FG02-04ER15571 is gratefully acknowledged.  
The computational resources and the support for Kochise Bennett was provided by DOE.
M.K gratefully acknowledges support from the Alexander von Humboldt foundation
through the Feodor Lynen program. We would like to thank the green planet cluster (NSF Grant CHE-0840513) for allocation of compute resources.
\end{acknowledgments}

\appendix

\section{Model parameters}
\label{app:params}
The model potentials shown in Fig. \ref{fig:pcat_bare}(a) are obtained from Morse potentials:
\begin{align}
V_i(q) = D_i\left[1 - \exp\left(-a_i(q-q_{0,i})\right) \right]^2 + V_0
\end{align}
The respective parameters are given in tab. \ref{tab:morse}.
\begin{table}
\caption{Parameters for the potentials shown in Fig. \ref{fig:pcat_bare}.\label{tab:morse}}
\begin{tabular}{lrrrr}
\hline\hline$i$ & $D_i$\,[eV] & $a_i$\,[\AA$^{-1}$] & $q_{0,i}$\,[\AA] & $V_0$\,[eV]\\\hline\hline
S$_0$ & 3.0   &  1     & 2.0 & 0\\
S$_1$ & 0.01 &  2.43& 2.5 & 3\\
S$_2$ & 3.0   &  1     & 2.3 & 4.5\\\hline\hline
\end{tabular}
\end{table}

The transition dipole shown in Fig. \ref{fig:pcat_bare}(b) is defined by the sigmoid function:
\begin{align}
\mu(q) = \dfrac{4}{1+\exp[2.4575 (q-4.232)]}
\end{align}

\section{Quantum Propagation}
\label{app:Propgation}
The wave packet propagations are carried out on a numerical grid using the Hamiltonian
from Eq. \ref{eq:Hkl_simple} where the kinetic energy is given by
\begin{align}
\hat T = -\dfrac{1}{2m} \dfrac{d^2}{dq^2}
\end{align}
with $m=3650$ being the reduced mass. 
For all dissociative potentials the kinetic
energy term $\hat{T}$ is replaced by a perfectly matched layer \cite{Nissen10}
to avoid spurious reflections at the edge of the grid.
The time evolution is calculated with an Arnoldi propagation scheme \cite{Tannor,Smyth98cpc}.

\section{Transient Absorption Spectrum}
\label{app:signal}
The transient absorption signal is linear in the probe intensity and given as the frequency integrated rate of change in the photon number (for further details see Ref. \cite{Dorfman15pra}):
\begin{align}
S_{N}(T) =& -\dfrac{2}{\hbar} {\cal I} \int_{-\infty}^{\infty}dt\int_{-\infty}^{t}d\tau
{\cal E}^*(t-T){\cal E}(\tau-T)\nonumber\\
& \times \br{\Psi_0} U^\dagger(t,0) \hat\mu U(t,\tau) \hat\mu^\dagger U(\tau,0)\ke{\Psi_0}
\end{align}
where $\Psi_0=(\hat\mu_{g+}+\hat\mu_{g-})\Psi_{S_0,v=0}$ is the initial wave function prepared by impulsive excitation from the vibrational ground state of the S$_0$ potential and
$U(t,t^\prime) = \exp\left(-i\hat H (t-t^\prime)\right)$ propagates the system from $t^\prime$ to $t$.
The propagator $U$ is implement by a numerical propagation as described in the previous section.
The operator $\hat\mu\equiv\mu_{ij}(q)$ includes transition dipole moments between the relevant electronic stats and depends on the nuclear coordinates. The probe field is given
by
\begin{align}
{\cal E}(t) = e^{-i\omega_Lt - t^2 /2\sigma^2}\,,
\end{align}
where $\omega_L$ is the center frequency and $\sigma$ the temporal width of the laser pulse
and $T$ is the delay with respect to the initial state preparation.

\end{document}